\documentclass[aps,pra,floatfix,showpacs,10pt,twocolumn]{revtex4}
\usepackage[dvips]{graphicx}
\usepackage{amsmath,amsfonts,amssymb,graphics,graphicx,epsfig,color,times,bbm}

\begin{document}

\title{Effect of different Dzyaloshinskii-Moriya interactions on entanglement in the Heisenberg XYZ chain}
\author{Da-Chuang Li$^{1,2}$\footnote{E-mail: dachuang@ahu.edu.cn}, Zhuo-Liang Cao$^{1,
2}$\footnote{E-mail:zhuoliangcao@gmail.com (Corresponding
Author)}}

\affiliation{$^{1}$Department of Physics, Hefei Teachers
College, Hefei 230061 P. R. China\\
$^{2}$School of Physics {\&} Material Science, Anhui University,
Hefei 230039 P. R. China}



\begin{abstract}
In this paper, we study the thermal entanglement in a two-qubit
Heisenberg XYZ system with different Dzyaloshinskii-Moriya (DM)
couplings. We show that different DM coupling parameters have
different influences on the entanglement and the critical
temperature. In addition, we find that when $J_{i}$ ($i$-component
spin coupling interaction) is the largest spin coupling
coefficient, $D_{i}$ ($i$-component DM interaction) is the most
efficient DM control parameter, which can be obtained by adjusting
the direction of DM interaction.
\end{abstract}

\maketitle

\section{introduction}
Entanglement has been studied intensely in recent years due to its
fascinating nonclassical feature and potential applications in
quantum information processing \cite{1}. As a simple system,
Heisenberg model is an ideal candidate for the generation and the
manipulation of entangled states. Many physical systems, such as
nuclear spins \cite{2}, quantum dots \cite{3}, superconductor
\cite{4} and optical lattices \cite{5}, have been simulated by
this model, and the Heisenberg interaction alone can be used for
quantum computation by suitable coding \cite{6}. Recently, the
Heisenberg models, including Ising model \cite{7}, XY model
\cite{8}, XXX model \cite{9}, XXZ model \cite{10} and XYZ model
\cite{11,12}, have been intensively studied. Shan \emph{et al.}
investigated the effects of DM interaction, impurity and exchange
couplings on entanglement in XY spin chain \cite{16}. Aydiner
\emph{et al.} studied the thermal entanglement of a two-qutrit
Ising system with DM interaction \cite{17}, they found that the
control of entanglement can be optimized by utilizing competing
effects of the magnetic field and the DM interaction. Wang
\emph{et al.} investigated the effects of the DM interaction and
intrinsic decoherence on entanglement teleportation in the
two-qubit XXX Heisenberg model \cite{18}.

In the above models, the influences of the z-component DM
interaction (arising from spin-orbit coupling) and the external
magnetic field on the entanglement have been discussed, but the DM
coupling interactions along other directions have never been taken
into account. Quite recently, we discussed the influences of
x-component DM interaction on entanglement in Heisenberg XXZ model
\cite{13} and XYZ model \cite{14}. To research further the
differences between DM coupling interactions along different
directions, in this paper, we generalize the special Heisenberg
models to the generalized Heisenberg XYZ models with different DM
interactions, and then analyze the different influences of $D_{x}$
(x-component DM control parameter), $D_{y}$ (y-component DM
control parameter) and $D_{z}$ (z-component DM control parameter)
on the entanglement and the critical temperature. We find that
$D_{i}$ is the most efficient DM control parameter when $J_{i}$ is
the largest spin coupling coefficient. Thus, according to the
relation among $J_{i}(i=x,y,z)$, we can know which is the most
efficient DM control parameter. In order to provide a detailed
analytical and numerical analysis, here we take concurrence as a
measure of entanglement \cite{15}. The concurrence $C$ ranges from
0 to 1, $C=0$ and $C=1$ indicate the vanishing entanglement and
the maximal entanglement respectively. For a mixed state $\rho$,
the concurrence of the state is
$C(\rho)=\max\{2\lambda_{\max}-\sum_{i=1}^{4}\lambda_{i}, 0\}$,
where $\lambda_{i}s$ are the positive square roots of the
eigenvalues of the matrix
$R=\rho(\sigma^{y}\bigotimes\sigma^{y})\rho^{*}(\sigma^{y}\bigotimes\sigma^{y})$,
and the asterisk denotes the complex conjugate.

This paper is organized as follows. In Sec. II, we introduce the
Heisenberg XYZ models with different DM interaction parameters,
and give the analytical expressions of the concurrences. In Sec.
III, we analyze the different influences of different DM control
parameters ($D_{x}$, $D_{y}$ and $D_{z}$) on the entanglement and
the critical temperature. Finally, in Sec. IV a discussion
concludes the paper.

\section{The Heisenberg XYZ models with different DM interaction parameters}
\subsection{Heisenberg XYZ model with $D_{x}$}
The Hamiltonian $H$ for a two-qubit anisotropic Heisenberg XYZ
chain with DM interaction parameter $D_{x}$ is
\begin{equation}
\label{1}
H=J_{x}\sigma_{1}^{x}\sigma_{2}^{x}+J_{y}\sigma_{1}^{y}\sigma_{2}^{y}+J_{z}\sigma_{1}^{z}\sigma_{2}^{z}
+D_{x}(\sigma_{1}^{y}\sigma_{2}^{z}-\sigma_{1}^{z}\sigma_{2}^{y}),
\end{equation}
where $J_{i}(i=x,y,z)$ are the real coupling coefficients, $D_{x}$
is the x-component DM coupling parameter, and
$\sigma^{i}(i=x,y,z)$ are the Pauli matrices. The coupling
constants $J_{i}>0$ corresponds to the antiferromagnetic case, and
$J_{i}<0$ corresponds to the ferromagnetic case. This model can be
reduced to some special Heisenberg models by changing $J_{i}$.
Parameters $J_{i}$ and $D_{x}$ are dimensionless.

In the standard basis
$\{|00\rangle,|01\rangle,|10\rangle,|11\rangle\}$, the Hamiltonian
(1) can be rewritten as
\begin{equation}
\label{2} H=\left(
\begin{array}{cccc}
  J_{z} & iD_{x} & -iD_{x} & J_{x}-J_{y} \\
  -iD_{x} & -J_{z} & J_{x}+J_{y} & iD_{x} \\
  iD_{x} & J_{x}+J_{y} & -J_{z} & -iD_{x} \\
  J_{x}-J_{y} & -iD_{x} & iD_{x} & J_{z} \\
\end{array}
\right).
\end{equation}
By calculating, we can obtain the eigenstates of $H$:
\begin{subequations}\label{3}
\begin{equation}
|\Psi_{1}\rangle=\frac{1}{\sqrt{2}}(|01\rangle+|10\rangle),
\end{equation}
\begin{equation}
|\Psi_{2}\rangle=\frac{1}{\sqrt{2}}(|00\rangle+|11\rangle),
\end{equation}
\begin{equation}
|\Psi_{3}\rangle=\frac{1}{\sqrt{2}}(\sin\theta_{1}|00\rangle-i\cos\theta_{1}|01\rangle+i\cos\theta_{1}|10\rangle-\sin\theta_{1}|11\rangle),
\end{equation}
\begin{equation}
|\Psi_{4}\rangle=\frac{1}{\sqrt{2}}(\sin\theta_{2}|00\rangle+i\cos\theta_{2}|01\rangle-i\cos\theta_{2}|10\rangle-\sin\theta_{2}|11\rangle),
\end{equation}
\end{subequations}
with corresponding eigenvalues:
\begin{subequations}\label{4}
\begin{equation}
E_{1,2}=J_{x}\pm J_{y}\mp J_{z},
\end{equation}
\begin{equation}
E_{3,4}=-J_{x}\pm w,
\end{equation}
\end{subequations}
where $\theta_{1,2}=\arctan(\frac{2D_{x}}{w\mp J_{y}\mp J_{z}})$,
and $w=\sqrt{4D^{2}_{x}+(J_{y}+J_{z})^{2}}$. The system state at
thermal equilibrium (thermal state) is
$\rho(T)=\frac{\exp(\frac{-H}{K_{B}T})}{Z}$, where
$Z=Tr[\exp(\frac{-H}{K_{B}T})]$ is the partition function of the
system, $H$ is the system Hamiltonian, $T$ is the temperature and
$K_{B}$ is the Boltzmann costant which we take equal to 1 for
simplicity. Thus, in the above standard basis, we can get the
following analytical expression of the density matrix $\rho(T)$:
\begin{equation}
\label{5} \rho(T)=\left(
\begin{array}{cccc}
  m_{1} & q & q^{*} & m_{2} \\
  q^{*} & n_{1} & n_{2} & q \\
  q & n_{2} & n_{1} & q^{*} \\
  m_{2} & q^{*} & q & m_{1} \\
\end{array}
\right),
\end{equation}
where
\begin{eqnarray*}
m_{1,2}=\frac{1}{2Z}\big(e^{-\frac{E_{2}}{T}}\pm
e^{-\frac{E_{3}}{T}}\sin^{2}\theta_{1}\pm
e^{-\frac{E_{4}}{T}}\sin^{2}\theta_{2}\big),
\end{eqnarray*}
\begin{eqnarray*}
n_{1,2}=\frac{1}{2Z}\big(e^{-\frac{E_{1}}{T}}\pm
e^{-\frac{E_{3}}{T}}\cos^{2}\theta_{1}\pm
e^{-\frac{E_{4}}{T}}\cos^{2}\theta_{2}\big),
\end{eqnarray*}
\begin{eqnarray*}
q=\frac{i}{2Z}\big(e^{-\frac{E_{3}}{T}}\sin\theta_{1}\cos\theta_{1}-e^{-\frac{E_{4}}{T}}\sin\theta_{2}\cos\theta_{2}\big).
\end{eqnarray*}
After straightforward calculations, the positive square roots of
the eigenvalues of the matrix
$R=\rho(\sigma^{y}\bigotimes\sigma^{y})\rho^{*}(\sigma^{y}\bigotimes\sigma^{y})$
can be expressed as:
\begin{subequations}\label{6}
\begin{eqnarray}
\lambda_{1,2}=\frac{1}{Z}e^{\frac{J_{x}\pm w}{T}},
\end{eqnarray}
\begin{eqnarray}
\lambda_{3,4}=\frac{1}{Z}e^{\frac{-J_{x}\pm J_{y}\mp J_{z}}{T}},
\end{eqnarray}
\end{subequations}
where
$Z=2e^{\frac{-J_{x}}{T}}\cosh(\frac{J_{y}-J_{z}}{T})+2e^{\frac{J_{x}}{T}}\cosh(\frac{w}{T})$.
Thus, the concurrence of this system can be expressed as
\cite{15}:
\begin{eqnarray}
\label{7} C=\left\{
\begin{array}{l}
\max\{|\lambda_{1}-\lambda_{3}|-\lambda_{2}-\lambda_{4},0\},\,\,\,if\,\,\ J_{y}>J_{z},\\
\max\{|\lambda_{1}-\lambda_{4}|-\lambda_{2}-\lambda_{3},0\},\,\,\,if\,\,\ J_{y}\leqslant J_{z}.\\
\end{array}
\right.
\end{eqnarray}
which is consistent with the results in Ref. \cite{13} for
$J_{x}=J_{y}$.

\subsection{Heisenberg XYZ model with $D_{y}$}
Here we consider the case of the two-qubit anisotropic Heisenberg
XYZ chain with y-component DM parameter $D_{y}$. The Hamiltonian
is
\begin{equation}
\label{8}
H^{\prime}=J_{x}\sigma_{1}^{x}\sigma_{2}^{x}+J_{y}\sigma_{1}^{y}\sigma_{2}^{y}+J_{z}\sigma_{1}^{z}\sigma_{2}^{z}
+D_{y}(\sigma_{1}^{z}\sigma_{2}^{x}-\sigma_{1}^{x}\sigma_{2}^{z}),
\end{equation}
where $D_{y}$ is the y-component DM coupling parameter, which is
also dimensionless.

In the standard basis
$\{|00\rangle,|01\rangle,|10\rangle,|11\rangle\}$, the Hamiltonian
(8) can be rewritten as
\begin{equation}
\label{9} H^{\prime}=\left(
\begin{array}{cccc}
  J_{z} & D_{y} & -D_{y} & J_{x}-J_{y} \\
  D_{y} & -J_{z} & J_{x}+J_{y} & D_{y} \\
  -D_{y} & J_{x}+J_{y} & -J_{z} & -D_{y} \\
  J_{x}-J_{y} & D_{y} & -D_{y} & J_{z} \\
\end{array}
\right).
\end{equation}
Similarly, by calculating, we can obtain the eigenstates of
$H^{\prime}$:
\begin{subequations}\label{10}
\begin{equation}
|\Psi_{1}^{\prime}\rangle=\frac{1}{\sqrt{2}}(|01\rangle+|10\rangle),
\end{equation}
\begin{equation}
|\Psi_{2}^{\prime}\rangle=\frac{1}{\sqrt{2}}(|00\rangle-|11\rangle),
\end{equation}
\begin{equation}
|\Psi_{3}^{\prime}\rangle=\frac{1}{\sqrt{2}}(\sin\phi_{1}|00\rangle-\cos\phi_{1}|01\rangle+\cos\phi_{1}|10\rangle+\sin\phi_{1}|11\rangle),
\end{equation}
\begin{equation}
|\Psi_{4}^{\prime}\rangle=\frac{1}{\sqrt{2}}(\sin\phi_{2}|00\rangle-\cos\phi_{2}|01\rangle+\cos\phi_{2}|10\rangle+\sin\phi_{2}|11\rangle),
\end{equation}
\end{subequations}
with corresponding eigenvalues:
\begin{subequations}\label{11}
\begin{equation}
E_{1,2}^{\prime}=J_{y}\pm J_{x}\mp J_{z},
\end{equation}
\begin{equation}
E_{3,4}^{\prime}=-J_{y}\pm w^{\prime},
\end{equation}
\end{subequations}
where $\phi_{1,2}=\arctan(\frac{2D_{y}}{J_{x}+J_{z}\mp
w^{\prime}})$, and
$w^{\prime}=\sqrt{4D^{2}_{y}+(J_{x}+J_{z})^{2}}$. In the above
standard basis, the density matrix $\rho^{\prime}(T)$ has the
following form:
\begin{equation}
\label{12} \rho^{\prime}(T)=\left(
\begin{array}{cccc}
  m_{1}^{\prime} & -q^{\prime} & q^{\prime} & m_{2}^{\prime} \\
  -q^{\prime} & n_{1}^{\prime} & n_{2}^{\prime} & -q^{\prime} \\
  q^{\prime} & n_{2}^{\prime} & n_{1}^{\prime} & q^{\prime} \\
  m_{2}^{\prime} & -q^{\prime} & q^{\prime} & m_{1}^{\prime} \\
\end{array}
\right),
\end{equation}
where
\begin{eqnarray*}
m_{1,2}^{\prime}=\frac{1}{2Z^{\prime}}\big(\pm
e^{-\frac{E_{2}^{\prime}}{T}}+
e^{-\frac{E_{3}^{\prime}}{T}}\sin^{2}\phi_{1}+
e^{-\frac{E_{4}^{\prime}}{T}}\sin^{2}\phi_{2}\big),
\end{eqnarray*}
\begin{eqnarray*}
n_{1,2}^{\prime}=\frac{1}{2Z^{\prime}}\big(e^{-\frac{E_{1}^{\prime}}{T}}\pm
e^{-\frac{E_{3}^{\prime}}{T}}\cos^{2}\phi_{1}\pm
e^{-\frac{E_{4}^{\prime}}{T}}\cos^{2}\phi_{2}\big),
\end{eqnarray*}
\begin{eqnarray*}
q^{\prime}=\frac{1}{2Z^{\prime}}\big(e^{-\frac{E_{3}^{\prime}}{T}}\sin\phi_{1}\cos\phi_{1}+e^{-\frac{E_{4}^{\prime}}{T}}\sin\phi_{2}\cos\phi_{2}\big).
\end{eqnarray*}
Then the positive square roots of the eigenvalues of the matrix
$R^{\prime}=\rho^{\prime}(\sigma^{y}\bigotimes\sigma^{y})\rho^{\prime
*}(\sigma^{y}\bigotimes\sigma^{y})$ can be obtained
\begin{subequations}\label{13}
\begin{eqnarray}
\lambda^{\prime}_{1,2}=\frac{1}{Z^{\prime}}e^{\frac{-J_{y}\pm
J_{x}\mp J_{z}}{T}},
\end{eqnarray}
\begin{eqnarray}
\lambda^{\prime}_{3,4}=\frac{1}{Z^{\prime}}e^{\frac{J_{y}\pm
w^{\prime}}{T}},
\end{eqnarray}
\end{subequations}
where
$Z^{\prime}=2e^{\frac{-J_{y}}{T}}\cosh(\frac{J_{x}-J_{z}}{T})+2e^{\frac{J_{y}}{T}}\cosh(\frac{w^{\prime}}{T})$.
Thus, the concurrence of this system can be expressed as:
\begin{eqnarray}
\label{14} C=\left\{
\begin{array}{l}
\max\{|\lambda^{\prime}_{1}-\lambda^{\prime}_{3}|-\lambda^{\prime}_{2}-\lambda^{\prime}_{4},0\},\,\,\,if\,\,\ J_{x}>J_{z},\\
\max\{|\lambda^{\prime}_{2}-\lambda^{\prime}_{3}|-\lambda^{\prime}_{1}-\lambda^{\prime}_{4},0\},\,\,\,if\,\,\ J_{x}\leqslant J_{z}.\\
\end{array}
\right.
\end{eqnarray}

\subsection{Heisenberg XYZ model with $D_{z}$}
The Hamiltonian $H^{\prime\prime}$ of a two-qubit anisotropic
Heisenberg XYZ chain with z-component DM parameter $D_{z}$ is
\begin{equation}
\label{15}
H^{\prime\prime}=J_{x}\sigma_{1}^{x}\sigma_{2}^{x}+J_{y}\sigma_{1}^{y}\sigma_{2}^{y}+J_{z}\sigma_{1}^{z}\sigma_{2}^{z}
+D_{z}(\sigma_{1}^{x}\sigma_{2}^{y}-\sigma_{1}^{y}\sigma_{2}^{x}),
\end{equation}
where $D_{z}$ is the z-component DM coupling parameter, which is
also dimensionless.

Using the similar process, we can get the eigenstates of
$H^{\prime\prime}$:
\begin{subequations}\label{16}
\begin{equation}
|\Psi_{1,2}^{\prime\prime}\rangle=\frac{1}{\sqrt{2}}(|00\rangle\pm|11\rangle),
\end{equation}
\begin{equation}
|\Psi_{3,4}^{\prime\prime}\rangle=\frac{1}{\sqrt{2}}(|01\rangle\pm\chi|10\rangle),
\end{equation}
\end{subequations}
with corresponding eigenvalues:
\begin{subequations}\label{17}
\begin{equation}
E_{1,2}^{\prime\prime}=J_{z}\pm J_{x}\mp J_{y},
\end{equation}
\begin{equation}
E_{3,4}^{\prime\prime}=-J_{z}\pm w^{\prime\prime},
\end{equation}
\end{subequations}
where
$\chi=\frac{J_{x}+J_{y}-2iD_{z}}{\sqrt{4D^{2}_{z}+(J_{x}+J_{y})^2}}$,
and $w^{\prime\prime}=\sqrt{4D^{2}_{z}+(J_{x}+J_{y})^{2}}$.
Similarly, we can get the analytical expressions of
$\rho(T)^{\prime\prime}$ and $R^{\prime\prime}$, but we do not
list them because of the complexity. After straightforward
calculations, the positive square roots of the eigenvalues of
$R^{\prime\prime}=\rho^{\prime\prime}(\sigma^{y}\bigotimes\sigma^{y})\rho^{\prime\prime
*}(\sigma^{y}\bigotimes\sigma^{y})$ can be expressed as:
\begin{subequations}\label{18}
\begin{eqnarray}
\lambda_{1,2}^{\prime\prime}=\frac{1}{Z^{\prime\prime}}
e^{\frac{J_{z}\pm w^{\prime\prime}}{T}},
\end{eqnarray}
\begin{eqnarray}
\lambda_{3,4}^{\prime\prime}=\frac{1}{Z^{\prime\prime}}e^{\frac{-J_{z}\pm
J_{x}\mp J_{y}}{T}},
\end{eqnarray}
\end{subequations}
with
$Z^{\prime\prime}=2e^{\frac{-J_{z}}{T}}\cosh(\frac{J_{x}-J_{y}}{T})+2e^{\frac{J_{z}}{T}}\cosh(\frac{w^{\prime\prime}}{T})$.
Thus, the concurrence of this system can be written as:
\begin{eqnarray}
\label{19} C=\left\{
\begin{array}{l}
\max\{|\lambda_{1}^{\prime\prime}-\lambda_{3}^{\prime\prime}|-\lambda_{2}^{\prime\prime}-\lambda_{4}^{\prime\prime},0\},\,\,\,if\,\,\ J_{x}>J_{y},\\
\max\{|\lambda_{1}^{\prime\prime}-\lambda_{4}^{\prime\prime}|-\lambda_{2}^{\prime\prime}-\lambda_{3}^{\prime\prime},0\},\,\,\,if\,\,\ J_{x}\leqslant J_{y}.\\
\end{array}
\right.
\end{eqnarray}
which is also consistent with the results in Ref. \cite{13} when
$J_{x}=J_{y}$.

\begin{figure}[tbp]
\includegraphics[scale=0.4,angle=0,bb=0 300 595 550]{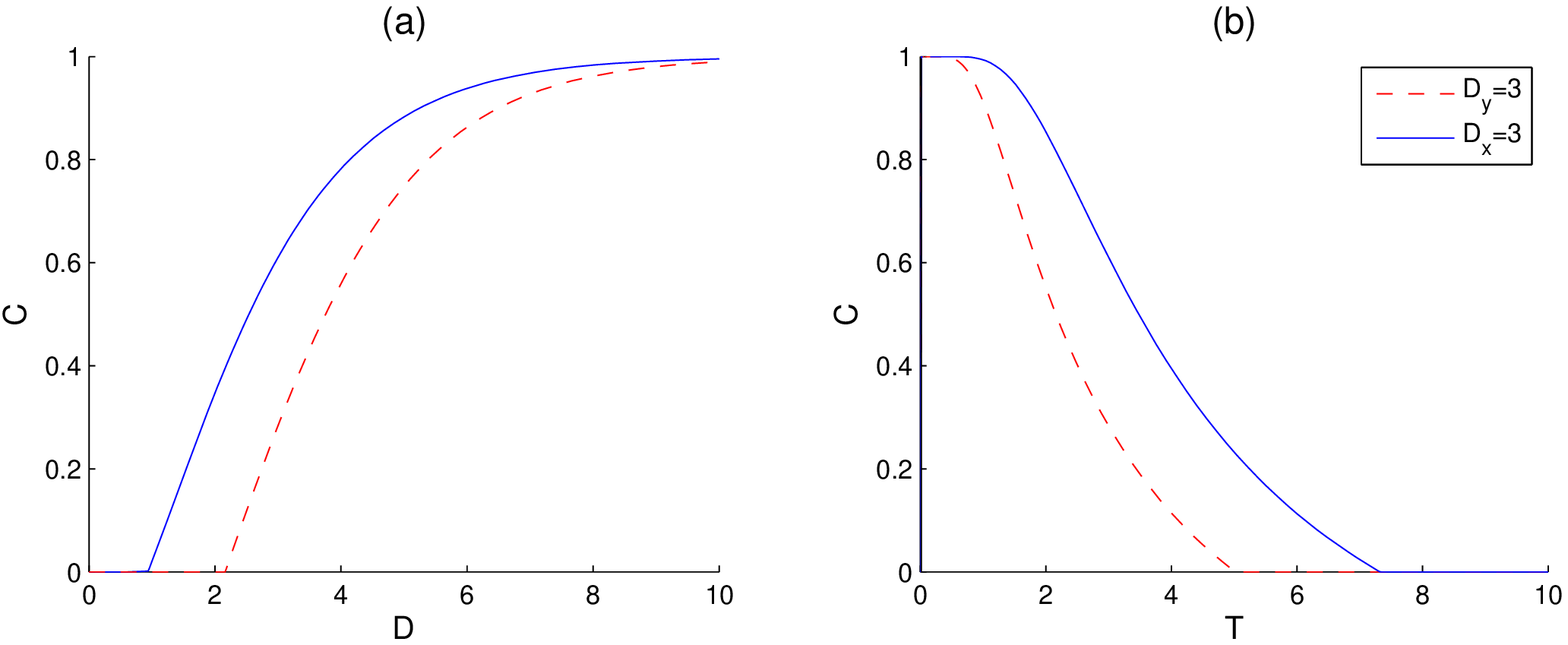}
\caption{(Color online) (a) The concurrence is plotted versus
$D_{x}$ (blue solid line) and $D_{y}$ (red dashed line) for $T=3$.
(b) The concurrence is plotted as a function of the temperature
$T$ for $D_{x}=3$ (blue solid line) and $D_{y}=3$ (red dashed
line). Here the coupling constants $J_{x}=0.2$, $J_{y}=-1$ and
$J_{z}=-0.5$.} \label{fig1}
\end{figure}

\begin{figure}[tbp]
\includegraphics[scale=0.4,angle=0,bb=0 295 595 550]{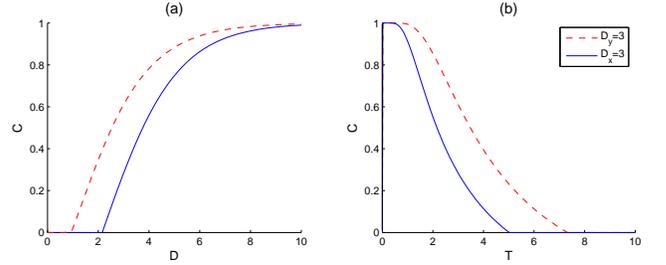}
\caption{(Color online) (a) The concurrence is plotted versus
$D_{x}$ (blue solid line) and $D_{y}$ (red dashed line) for $T=3$.
(b) The concurrence is plotted as a function of the temperature
$T$ for $D_{x}=3$ (blue solid line) and $D_{y}=3$ (red dashed
line). Here the coupling constants $J_{x}=-1$, $J_{y}=0.2$ and
$J_{z}=-0.5$.} \label{fig2}
\end{figure}

From Eqs. (7), (14) and (19), one can see that when $J_{x}=J_{y}$,
there is the same entanglement for $D_{x}=D_{y}$; when
$J_{y}=J_{z}$, there is the same entanglement for $D_{y}=D_{z}$;
and when $J_{x}=J_{z}$, there is the same entanglement for
$D_{x}=D_{z}$. So when $J_{x}=J_{y}=J_{z}$, there is also the same
entanglement for the same values of DM interaction parameters
($D_{x}$, $D_{y}$ and $D_{z}$)

\section{The comparison between the different DM control parameters in Heisenberg XYZ model}
\subsection{The comparison between $D_{x}$ and $D_{y}$}

In Heisenberg XYZ model, we has analyzed all kinds of spin
coupling coefficients satisfying $J_{x}>J_{y}$. For simplicity,
here we choose one set of spin coupling coefficients satisfying
$J_{x}>J_{y}$, and plot Fig. 1 to demonstrate the properties of
different DM parameters. In Fig. 1(a), we find the entanglement
increases with the increase of DM coupling parameter. Furthermore,
the critical value of $D_{x}$ is smaller than $D_{y}$, and $D_{x}$
has more entanglement for $D_{x}=D_{y}$. In Fig. 1(b), we find
that increasing temperature will decrease the entanglement, and
$D_{x}$ has a higher critical temperature than the same $D_{y}$,
so that the entanglement can exist at higher temperatures for
$D_{x}$.

Similarly, we has analyzed various spin coupling coefficients
satisfying $J_{x}<J_{y}$. For simplicity, we choose one set of
coupling coefficients satisfying $J_{x}<J_{y}$, and plot Fig. 2 to
demonstrate the properties of different DM parameters. In Fig.
2(a), it is shown that increasing the DM coupling parameter can
enhance the entanglement. Besides, for a certain temperature, the
critical value of $D_{y}$ is smaller than $D_{x}$, and $D_{y}$ has
more entanglement for $D_{x}=D_{y}$. In Fig. 2(b), it is easy to
see that $D_{y}$ has a higher critical temperature than the same
$D_{x}$, so that the entanglement can exist at higher temperatures
for $D_{y}$.

Thus, the x-component parameter $D_{x}$ has a smaller critical
value, higher critical temperature and more entanglement than the
same y-component parameter $D_{y}$ for $J_{x}>J_{y}$, and $D_{y}$
has a smaller critical value, higher critical temperature and more
entanglement than the same $D_{x}$ for $J_{x}<J_{y}$.

\subsection{The comparison between $D_{y}$ and $D_{z}$}

Similarly, for $J_{y}>J_{z}$ case, Fig. 3 is plotted to show the
properties of different DM parameters in Heisenberg XYZ model. In
Fig. 3(a), we can see that the y-component parameter $D_{y}$ has a
smaller critical value, and more entanglement for $D_{y}=D_{z}$.
In Fig. 3(b), we can see that $D_{y}$ has a higher critical
temperature than the same $D_{z}$, so that the entanglement can
exist at higher temperatures for $D_{y}$.
\begin{figure}[tbp]
\includegraphics[scale=0.4,angle=0,bb=0 295 595 550]{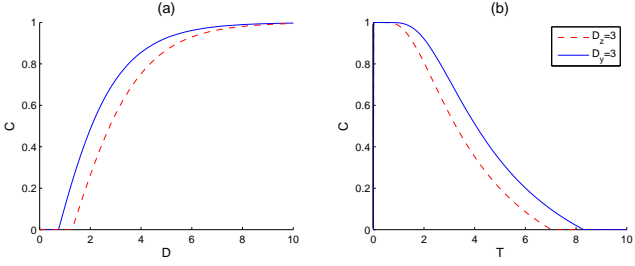}
\caption{(Color online) (a) The concurrence is plotted versus
$D_{y}$ (blue solid line) and $D_{z}$ (red dashed line) for $T=3$.
(b) The concurrence is plotted as a function of the temperature
$T$ for $D_{y}=3$ (blue solid line) and $D_{z}=3$ (red dashed
line). Here the coupling constants $J_{x}=-0.5$, $J_{y}=1$ and
$J_{z}=0.2$.} \label{fig3}
\end{figure}

\begin{figure}[tbp]
\includegraphics[scale=0.4,angle=0,bb=0 295 595 550]{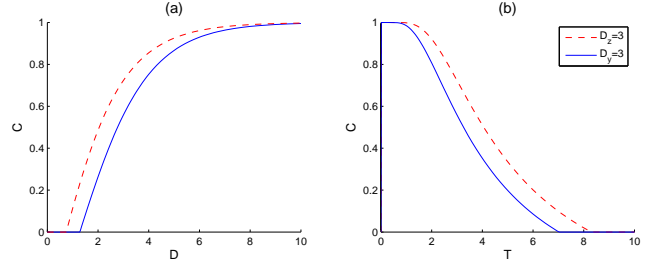}
\caption{(Color online) (a) The concurrence is plotted versus
$D_{y}$ (blue solid line) and $D_{z}$ (red dashed line) for $T=3$.
(b) The concurrence is plotted as a function of the temperature
$T$ for $D_{y}=3$ (blue solid line) and $D_{z}=3$ (red dashed
line). Here the coupling constants $J_{x}=-0.5$, $J_{y}=0.2$ and
$J_{z}=1$.} \label{fig4}
\end{figure}
Contrarily, for $J_{y}<J_{z}$ case, the concurrence versus
different parameters is shown in Fig. 4. In Fig. 4(a), it is easy
to see that the z-component parameter $D_{z}$ has a smaller
critical value, and more entanglement for $D_{y}=D_{z}$. In Fig.
4(b), it is easy to see that $D_{z}$ has a higher critical
temperature than the same $D_{y}$, so that the entanglement can
also exist at higher temperatures for $D_{z}$.

Thus, the y-component parameter $D_{y}$ has a smaller critical
value, higher critical temperature and more entanglement than the
same z-component parameter $D_{z}$ for $J_{y}>J_{z}$, and $D_{z}$
has a smaller critical value, higher critical temperature and more
entanglement than the same $D_{y}$ for $J_{y}<J_{z}$.

\subsection{The comparison between $D_{x}$ and $D_{z}$}

Here, for $J_{x}>J_{z}$ case, we plot Fig. 5 to illustrate the
properties of different DM parameters in Heisenberg XYZ model. In
Fig. 5(a), $D_{x}$ has a smaller critical value and more
entanglement for $D_{x}=D_{z}$. In Fig. 5(b), $D_{x}$ has a higher
critical temperature than the same $D_{z}$, so that the
entanglement can exist at higher temperatures for $D_{x}$.

For $J_{x}<J_{z}$ case, we plot Fig. 6 to illustrate the
properties of different DM parameters. In Fig. 6(a), for a certain
temperature, $D_{z}$ has a smaller critical value and more
entanglement for $D_{x}=D_{z}$. In Fig. 6(b), $D_{z}$ has a higher
critical temperature than the same $D_{x}$, so that the
entanglement can also exist at higher temperatures for $D_{z}$.

Thus, the x-component parameter $D_{x}$ has a smaller critical
value, higher critical temperature and more entanglement than the
same z-component parameter $D_{z}$ for $J_{x}>J_{z}$, and $D_{z}$
has a smaller critical value, higher critical temperature and more
entanglement than the same $D_{x}$ for $J_{x}<J_{z}$.

The above results indicate that when $J_{i}$ is the largest spin
coupling coefficient, the i-component DM control parameter $D_{i}$
has the smallest critical value, the highest critical temperature
and the most entanglement. So according to the relation among spin
coupling coefficients, we can know which is the most efficient DM
control parameter.
\begin{figure}[tbp]
\includegraphics[scale=0.4,angle=0,bb=0 295 595 550]{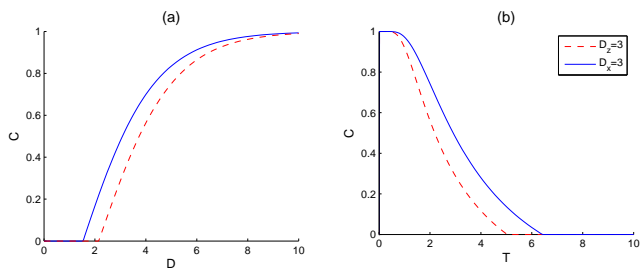}
\caption{(Color online) (a) The concurrence is plotted versus
$D_{x}$ (blue solid line) and $D_{z}$ (red dashed line) for $T=3$.
(b) The concurrence is plotted as a function of the temperature
$T$ for $D_{x}=3$ (blue solid line) and $D_{z}=3$ (red dashed
line). Here the coupling constants $J_{x}=-0.2$, $J_{y}=0.3$ and
$J_{z}=-1$.} \label{fig5}
\end{figure}

\begin{figure}[tbp]
\includegraphics[scale=0.4,angle=0,bb=0 295 595 550]{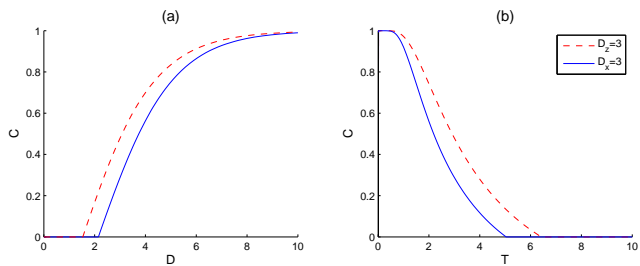}
\caption{(Color online) (a) The concurrence is plotted versus
$D_{x}$ (blue solid line) and $D_{z}$ (red dashed line) for $T=3$.
(b) The concurrence is plotted as a function of the temperature
$T$ for $D_{x}=3$ (blue solid line) and $D_{z}=3$ (red dashed
line). Here the coupling constants $J_{x}=-1$, $J_{y}=0.3$ and
$J_{z}=-0.2$.} \label{fig6}
\end{figure}

\section{Discussion}
We have investigated the thermal entanglement in a two-qubit
Heisenberg XYZ system with different Dzyaloshinskii-Moriya (DM)
couplings. We find that different DM interaction parameters
($D_{x}$, $D_{y}$ and $D_{z}$) have different influences on the
entanglement and the critical temperature. When $J_{i}(i=x,y,z)$
is the largest spin coupling coefficient, the $i$-component DM
interaction $D_{i}(i=x,y,z)$ has the smallest critical value, the
highest critical temperature and the most entanglement. In
addition, when $J_{x}=J_{y}=J_{z}$, there is the same entanglement
for the same values of DM interaction parameters ($D_{x}$, $D_{y}$
and $D_{z}$). Thus, according to the relation among spin coupling
coefficients ($J_{x}$, $J_{y}$ and $J_{z}$), the most efficient DM
control parameter can be obtained by adjusting the direction of DM
interaction.

\begin{acknowledgments}
This work is supported by National Natural Science Foundation of
China (NSFC) under Grant Nos: 60678022 and 10704001, the
Specialized Research Fund for the Doctoral Program of Higher
Education under Grant No. 20060357008, Anhui Provincial Natural
Science Foundation under Grant No. 070412060, and the Key Program
of the Education Department of Anhui Province under Grant No.
KJ2008A28ZC.
\end{acknowledgments}


\begin{thebibliography}{99}
\bibitem{1} Phys. World \textbf{11}, 33 (1998), special issue on quantum information; C. H. Bennett and S. J. Wiesner, Phys. Rev. Lett. \textbf{69}, 2881 (1992);
            C. H. Bennett \emph{et al.}, \emph{ibid.} \textbf{70}, 1895 (1993); A. K. Ekert, \emph{ibid.} \textbf{67}, 661 (1991); M. Murao \emph{et al.}, Phys. Rev. A \textbf{59}, 156 (1999).
\bibitem{2} B. E. Kane, Nature (London) \textbf{393}, 133 (1998).
\bibitem{3} D. Loss and D. P. DiVincenzo, Phys. Rev. A \textbf{57}, 120 (1998); G. Burkard, D. Loss, and D. P. DiVincenzo, Phys. Rev. B \textbf{59}, 2070
            (1999); B. Trauzettel \emph{et al.}, Nature Phys. \textbf{3} 192 (2007).
\bibitem{4} T. Senthil \emph{et al.}, Phys. Rev. B \textbf{60} 4245 (1999); M. Nishiyama \emph{et al.}, Phys. Rev. Lett. \textbf{98} 047002 (2007).
\bibitem{5} A. S\o rensen and K. M\o lmer, Phys. Rev. Lett. \textbf{83}, 2274 (1999).
\bibitem{6} D. A. Lidar, D. Bacon, and K. B. Whaley, Phys. Rev. Lett. \textbf{82}, 4556 (1999); D. P. DiVincenzo \emph{et al.}, Nature (London) \textbf{408}, 339 (2000); L. F. Santos, Phys. Rev. A \textbf{67}, 062306 (2003).
\bibitem{7} D. Gunlycke \emph{et al.}, Phys. Rev. A \textbf{64}, 042302 (2001); Z. Yang \emph{et al.}, Phys. Rev. Lett. \textbf{100}, 067203 (2008).
\bibitem{8} G. L. Kamta and Anthony F. Starace, Phys. Rev. Lett. \textbf{88}, 107901 (2002); X. G. Wang, Phys. Rev. A \textbf{66}, 034302 (2002); Y. Sun \emph{et al.}, Phys. Rev. A \textbf{68}, 044301 (2003);
            Z. C. Kao \emph{et al.}, Phys. Rev. A \textbf{72}, 062302 (2005); S. L. Zhu, Phys. Rev. Lett. \textbf{96}, 077206 (2006).
\bibitem{9} M. Asoudeh and V. Karimipour, Phys. Rev. A \textbf{71}, 022308 (2005); G. F. Zhang, Phys. Rev. A \textbf{75}, 034304 (2007).
\bibitem{10} G. F. Zhang and S. S. Li, Phys. Rev. A \textbf{72}, 034302 (2005); Y. Zhou \emph{et al.}, Phys. Rev. A \textbf{75}, 062304 (2007);
             M. Kargarian \emph{et al.}, Phys. Rev. A \textbf{77}, 032346 (2008).
\bibitem{11} D. C. Li and Z. L. Cao, Optics Communications, in press; L. Zhou \emph{et al.}, Phys. Rev. A \textbf{68}, 024301 (2003); G. H. Yang \emph{et al.}, e-print arXiv:quant-ph/0602051.
\bibitem{12} F. Kheirandish \emph{et al.}, Phys. Rev. A \textbf{77}, 042309 (2008); Z. N. Gurkan and O. K. Pashaev, e-print arXiv:quant-ph/0705.0679 and
             arXiv:quant-ph/0804.0710.
\bibitem{16} Chuan-Jia Shan \emph{et al.}, Chin. Phys. Lett. \textbf{25}, 817 (2008).
\bibitem{17} C. Aky\"{u}z, E. Aydiner, and \"{O}. E. M\"{u}stecaplio\v{g}lu, Optics Communications \textbf{281}, 5271 (2008).
\bibitem{18} Liang Qiu, An Min Wang, and Xiao Qiang Su, Physica A \textbf{387}, 6686 (2008).
\bibitem{13} D. C. Li, X. P. Wang, and Z. L. Cao, J. Phys. Condens. Matter \textbf{20} 325229 (2008).
\bibitem{14} D. C. Li and Z. L. Cao, Eur. Phys. J. D \textbf{50}, 207 (2008).
\bibitem{15} S. Hill and W. K. Wootters, Phys. Rev. Lett. \textbf{78}, 5022 (1997); W. K. Wootters, Phys. Rev. Lett. \textbf{80}, 2245 (1998).
\end{thebibliography}
\end{document}